\documentclass[aps,pre,reprint,superscriptaddress]{revtex4-2}
\usepackage{lineno,hyperref}
\usepackage{amsfonts}
\usepackage{amssymb}
\usepackage{amsmath}
\usepackage{graphicx}
\usepackage{makeidx}
\usepackage{subfig}
\usepackage{tabularx}
\usepackage{dcolumn}
\usepackage{soul}

\begin{document}

	\title{Dynamics and stabilization of topological edge solitons in driven–damped nonlinear SSH lattices}
	
	\author{A.\ Yosia}
	\author{R.\ Rusin}
	\email{rahmirus@sci.ui.ac.id}
	
	\affiliation{Department of Mathematics, Faculty of Mathematics and Natural Sciences, Universitas Indonesia,\\ Gedung D Lt.\ 2 FMIPA Kampus UI Depok, 16424, Indonesia}
	\affiliation{Nonlinear Dynamics Research Group, Universitas Indonesia,\\ Gedung D Lt.\ 2 FMIPA Kampus UI Depok, 16424, Indonesia}
	
	\author{R.\ Kusdiantara}
	
	\affiliation{Industrial and Financial Mathematics Research Group, Institut Teknologi Bandung,\\ Jl.\ Ganesha No.\ 10, Bandung, 40132, Indonesia}
	\affiliation{Centre of Mathematical Modelling and Simulation, Institut Teknologi Bandung,\\ Jl.\ Ganesha No.\ 10, Bandung, 40132, Indonesia}
	
	\author{H. Susanto}
	
	\affiliation{Department of Mathematics, Khalifa University, Abu Dhabi Campus, PO Box
		127788,\\ Abu Dhabi, United Arab Emirates}
	
	

	\date{\today}
	
\begin{abstract}
We study topological edge solitons in a nonlinear Su--Schrieffer--Heeger (SSH) lattice subject to parametric driving and linear damping. Starting from a vertically driven pendulum chain, we derive an effective driven--damped nonlinear SSH model and investigate its stationary edge-localized states. Analytical calculations reveal the existence of two phase-locked dissipative edge-soliton families that emerge from the nonlinear continuation of the topological edge mode. Using numerical continuation and spectral stability analysis, we construct the corresponding nonlinear branches and determine their stability properties. We show that parametric driving and damping fundamentally modify the conservative edge-state family by generating two dissipative branches with markedly different stability characteristics: one branch remains predominantly unstable, whereas the other develops substantially larger stability regions and significantly weaker instability growth rates. Direct numerical simulations further demonstrate that the robust branch can remain strongly localized over long time intervals even when weakly unstable. Simulations of the full driven--damped Klein--Gordon pendulum chain confirm the persistence of the edge-localized dynamics predicted by the reduced model. These results identify parametric driving and damping as an effective mechanism for enhancing the robustness and persistence of nonlinear topological localization in active lattice systems.
\end{abstract}

	\maketitle
	
	
	\section{Introduction}\label{sec:level1}
	
	Discrete solitons are spatially localized nonlinear excitations that arise in lattices with intrinsic discreteness and nonlinearity. They have been extensively studied in a variety of physical contexts, including optical waveguide arrays, Bose--Einstein condensates in optical lattices, micromechanical resonator chains, and electrical transmission lines~\cite{kevrekidis2009discrete, flach2008discrete, christodoulides2003discretizing, morsch2006dynamics, english2014experimental}. In such systems, the interplay between nonlinearity and dispersion (or inter-site coupling) enables the formation of stationary or traveling localized modes whose energy remains confined to a few lattice sites. The discrete nonlinear Schr\"odinger (DNLS) equation and its generalizations provide canonical models for describing these phenomena~\cite{kevrekidis2009discrete, flach2008discrete}. Beyond their fundamental significance, discrete solitons play an important role in nonlinear energy transport, signal control, and the design of reconfigurable metamaterials~\cite{lederer2008discrete}.
	
	Topologically protected states, on the other hand, emerge from nontrivial global properties of the underlying band structure~\cite{hasan2010colloquium, qi2011topological, asboth2016short,parker2024topological}. A paradigmatic example is the Su--Schrieffer--Heeger (SSH) lattice, which consists of a one-dimensional chain with alternating intersite couplings~\cite{su1979solitons}. In the linear regime, this system supports localized edge modes whose existence is guaranteed by chiral symmetry and characterized by a quantized topological invariant. These modes are robust against moderate disorder and perturbations that do not close the bulk spectral gap~\cite{hasan2010colloquium, qi2011topological}. The SSH model and its extensions have spurred extensive research across condensed-matter physics, photonics, and mechanical metamaterials~\cite{ozawa2019topological, huber2016topological}.

	When nonlinearity is introduced into topological lattices, the continuation of linear edge modes gives rise to nonlinear localized states, often referred to as edge solitons~\cite{lumer2013self, leykam2016edge, smirnova2020nonlinear,johansson2023topological}. In contrast to the linear case, the existence and robustness of these states are no longer guaranteed by a global spectral invariant, and their stability becomes strongly dependent on system parameters. In nonlinear SSH-type models, previous studies have shown that edge solitons can become unstable over significant regions of parameter space due to interactions with extended modes and other resonance mechanisms~\cite{ablowitz2014linear, hadad2016self, leykam2016edge,ma2021topological,sone2025transition,bugarski2025edge,chaunsali2021stability}. Understanding and controlling such instabilities is therefore crucial for realizing robust nonlinear topological excitations. 
	
	In conservative nonlinear SSH lattices, the stability of edge solitons is highly parameter dependent and may be lost through interactions with bulk modes or internal resonances. 
    Here we show that parametric driving and damping do not merely suppress or enhance stability uniformly. Instead, they reorganize the stability landscape and create selective stability windows. This driven--dissipative balance provides a controllable mechanism for stabilizing nonlinear topological states that is absent in the conservative SSH lattice.
	
	Parametric driving and damping have long been recognized as effective mechanisms for controlling the dynamics of nonlinear lattice systems~\cite{barashenkov1991stability,barashenkov1996existence}. In driven--damped variants of the discrete nonlinear Schr\"odinger (DNLS) equation, these effects can balance energy exchange between localized and extended modes, leading to the formation of stable dissipative solitons and breathers~\cite{susanto2006stability, syafwan2010discrete, syafwan2012solitons, cabanas2021dissipative,carreno2025oscillatory}. In such systems, parametric driving injects energy into the lattice, while damping removes excess energy, and their interplay can stabilize stationary states that are unstable in the conservative limit. This balance between energy injection and dissipation enables robust localized structures in a variety of physical settings, including optical cavities, mechanical resonator arrays, and nonlinear metamaterials.
	
	In this work, we investigate the stabilization of topological edge solitons in a parametrically driven and damped nonlinear SSH lattice. We show that the nonlinear continuation of the topological edge mode, which is often unstable in the conservative setting, can be stabilized through the combined action of parametric drive and damping. Using numerical continuation, we compute stationary solutions and construct bifurcation and stability diagrams as functions of the detuning, coupling contrast, and drive-damping parameters. Linear stability analysis reveals the emergence of stable edge solitons in parameter regimes where dissipative effects suppress instabilities induced by nonlinear mode interactions. Our results identify a general mechanism by which driven--dissipative effects can stabilize nonlinear topological states, thereby extending the range of parameters over which robust edge localization can be achieve
    d.

    The contribution of this work is fivefold. First, we derive a parametrically driven--damped nonlinear SSH-DNLS equation as a slow-envelope model for a vertically driven pendulum chain. Second, we analyze the linear essential and edge spectra and determine the onset condition for stationary edge-localized states. Third, we obtain an analytical approximation for phase-locked dissipative edge solitons, which reveals how driving and damping deform the conservative nonlinear edge branch. Fourth, using numerical continuation and spectral stability analysis, we identify selective stability windows generated by the balance between parametric gain and damping. Finally, we assess the validity of the reduced model through direct comparisons with simulations of the original pendulum chain.
	
	\section{Model formulation}\label{sec:model}
	
	\begin{figure}[tbhp!]
		\centering
		{\includegraphics[scale=0.4]{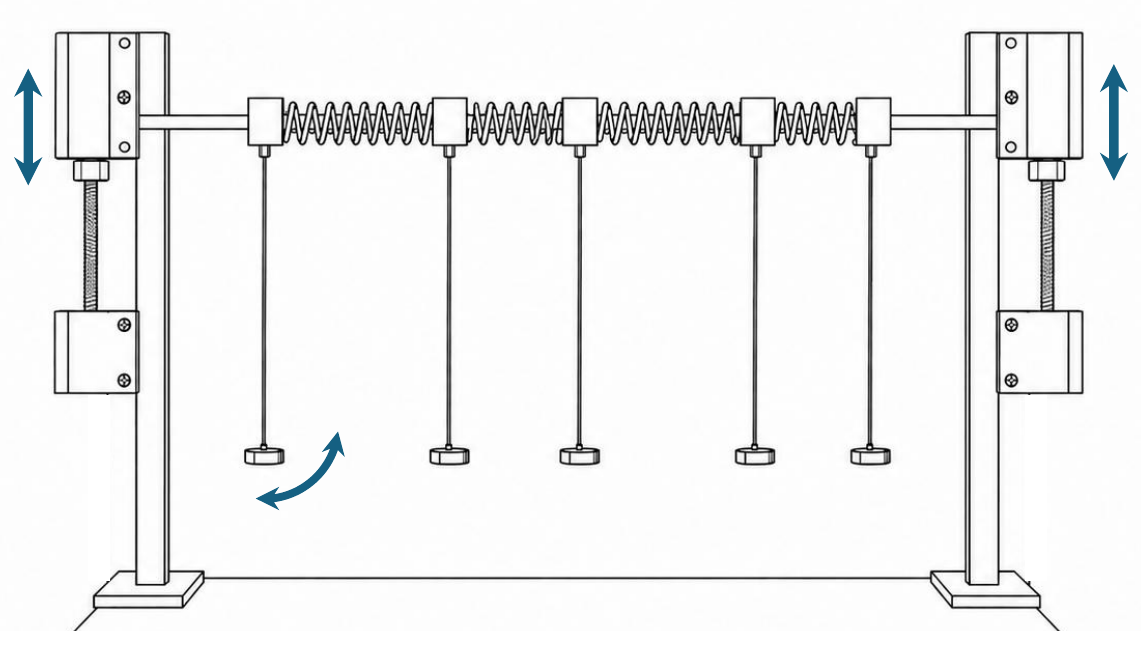}}
		\caption{Sketch of a possible experimental setup for a parametrically driven pendulum chain. Identical pendula are coupled via torsional springs mounted on a common support that oscillates vertically with a prescribed harmonic motion, thereby providing parametric excitation. Alternating coupling stiffnesses (indicated by varying spring lengths) emulate the dimerized Su--Schrieffer--Heeger (SSH) configuration. }
		\label{fig:sketch}
	\end{figure}
	
	We consider a one-dimensional chain of \(N\) identical pendula of mass \(m\) and length \(l\), coupled through torsional springs with alternating stiffnesses, forming a dimerized mechanical lattice analogous to the Su--Schrieffer--Heeger (SSH) model (see Fig.~\ref{fig:sketch}). The coupling stiffness between pendula \(n\) and \(n{+}1\) is given by
	\[
	K_n =
	\begin{cases}
		K_1, & n \text{ odd},\\[2pt]
		K_2, & n \text{ even},
	\end{cases}
	\qquad K_1, K_2 > 0,
	\]
	so that the lattice consists of alternating strong and weak couplings. The angular displacement of the \(n\)-th pendulum from the vertical is denoted by \(u_n(t)\).
	
	The common support board oscillates vertically according to
	\[
	y(t)=a\cos(\Omega t),
	\]
	which induces a parametric modulation of the effective gravitational acceleration, since \(\ddot{y}(t)=-a\,\Omega^2\cos(\Omega t)\), i.e., Kapitza's pendulum \cite{kapitza1951dynamic,kapitza1951pendulum}. Including linear viscous damping with coefficient \(\gamma\), the equations of motion take the form
	\begin{align}
		m l^2 \ddot{u}_n &+ \gamma \dot{u}_n
		+ m l \big(g+\ddot{y}(t)\big) \sin u_n \nonumber\\
		&= K_{n-1}(u_{n-1}-u_n) + K_n (u_{n+1}-u_n).
		\label{eq:ssh-full}
	\end{align}
	
	For the analytical treatment, we consider a semi-infinite lattice with boundary condition \(u_0 = 0\). In numerical simulations, a finite chain is used with an additional boundary condition imposed at the opposite end.
	
	Introducing the natural frequency \(\Omega_0^2=g/l\) and rescaling time by \(\tilde{t}=\Omega_0 t\), Eq.~\eqref{eq:ssh-full} becomes
	\begin{align}
		u_n'' + \epsilon^2\Gamma \,u_n'
		&+ \left[1-2\epsilon^2h\,\cos\left(\frac{\Omega}{\Omega_0}\tilde{t}\right)\right]\sin(u_n)
		\nonumber \\
		&=
		\epsilon^2c_{n-1}(u_{n-1}-u_n)+\epsilon^2c_n(u_{n+1}-u_n),
		\label{eq:nondim}
	\end{align}
	where primes denote derivatives with respect to \(\tilde{t}\), and the dimensionless parameters are defined as
	\[
	\epsilon^2\Gamma=\frac{\gamma}{ml^2\Omega_0},
	\qquad
	2\epsilon^2h=\frac{a\Omega^2}{l\Omega_0^2},
	\qquad
	\epsilon^2c_n=\frac{K_n}{ml^2\Omega_0^2}.
	\]
	
	We assume weak damping, parametric forcing, and coupling by considering \(0<\epsilon\ll1\), and restrict attention to small-amplitude oscillations. Expanding the sine nonlinearity,
	\[
	\sin u_n = u_n - \frac{u_n^3}{6} + \mathcal{O}(u_n^5),
	\]
	we focus on the regime of 2:1 parametric resonance by setting
	\[
	\frac{\Omega}{\Omega_0} = 2 - \epsilon^2\Delta,
	\]
	where the detuning is chosen as
	\begin{equation}
	    \Delta=\omega-c_n-c_{n-1}.
        \label{eq:detuning}
	\end{equation}
	Since $c_n +c_{n-1}=c_1+c_2$ in the bulk, the above choice corresponds to a constant effective detuning after the on-site contribution of the discrete Laplacian has been absorbed into $\omega$.  Boundary effects are then incorporated through the imposed edge condition. Later, after the phase transformation explained below, the reduced equation takes the standard SSH-DNLS form with coupling only through nearest-neighbor terms.
	
	To describe the slow modulation of the oscillations, we introduce the multiple time scales \(\tau=\tilde{t}\) and \(T=\epsilon^2\tilde{t}\). At leading order, the dynamics are governed by
	\[
	\partial_\tau^2 u_n + u_n = 0,
	\]
	with solution
	\begin{equation}
		u_n(\tau,T)=\sqrt{2}\,\epsilon\left(\Psi_n(T)e^{i\tau}+\Psi_n^*(T)e^{-i\tau}\right).
	\end{equation}
	
	Eliminating secular terms at order \(\mathcal{O}(\epsilon^3)\) produces the envelope equation
	\begin{align}
		i\dot{\Psi}_n
		&	+i\Gamma\Psi_n
		-h\Psi_n^*e^{-2i\Delta T}-|\Psi_n|^2\Psi_n=\nonumber\\
		&    c_{n-1}(\Psi_{n-1}-\Psi_n)
		+c_n(\Psi_{n+1}-\Psi_n),
		\label{eq:envelope_pre}
	\end{align}
	where \(\dot{\Psi}_n=d\Psi_n/dT\). 
	
	Substituting the phase transformation \(\Psi_n=\psi_n\exp(-i\Delta T)\) into Eq.~\eqref{eq:envelope_pre}, multiplying by $\exp(i\Delta T)$ and using \eqref{eq:detuning}, we obtain
	\begin{align}
		i\dot{\psi}_n
		=
		&	c_{n-1}\psi_{n-1}
		+c_n\psi_{n+1}\nonumber\\
		&-\left(\omega+i\Gamma\right)\psi_n+|\psi_n|^2\psi_n+h\,\psi_n^*.
		\label{eq:ssh-pddnls}
	\end{align}
	
	Equation~\eqref{eq:ssh-pddnls} is the parametrically driven and damped nonlinear Su--Schrieffer--Heeger (SSH) lattice, obtained as the amplitude equation governing the slow dynamics of the vertically driven pendulum chain. The derivation of the discrete nonlinear Schr\"odinger equation~\eqref{eq:ssh-pddnls} from the parametrically driven and damped nonlinear Klein--Gordon equation~\eqref{eq:ssh-full} can be rigorously justified following Ref.~\cite{muda2019justification} by establishing error bounds using energy estimates. In Section \ref{sec:numerics}, this reduction is further assessed by comparing the reduced SSH dynamics \eqref{eq:ssh-pddnls} with direct simulations of \eqref{eq:ssh-full}.
	
	In the absence of damping and drive (\(\Gamma = h = 0\)), Eq.~\eqref{eq:ssh-pddnls} reduces to the nonlinear SSH model, which supports nonlinear continuations of topologically protected edge states that are generally unstable in wide parameter regions \cite{ma2021topological}. The inclusion of damping and parametric forcing introduces a mechanism for stabilizing such localized edge solitons, as demonstrated in this work.

\section{Stationary Edge States and Their Spectral Properties}
\label{sec:analysis}
To exploit the dimerized structure of the lattice, we group two neighboring
sites into a single unit cell and introduce
\begin{equation}
\psi_{2n-1}(T)=a_n(T),
\qquad
\psi_{2n}(T)=b_n(T),
\qquad
n\in\mathbb{N}.
\label{eq:ab_variables}
\end{equation}
The alternating couplings are written as
\begin{equation}
c_{2n-1}=c_1,
\qquad
c_{2n}=c_2.
\label{eq:alternating_couplings}
\end{equation}
We consider a semi-infinite lattice with a left boundary imposed through
\begin{equation}
b_0=0.
\label{eq:left_boundary_ab}
\end{equation}
In these variables, Eq.~\eqref{eq:ssh-pddnls} becomes
\begin{equation}
\begin{aligned}
i\dot a_n
&=
c_1 b_n+c_2 b_{n-1}
+|a_n|^2a_n
-(\omega+i\Gamma)a_n
+h a_n^*,\\[4pt]
i\dot b_n
&=
c_1 a_n+c_2 a_{n+1}
+|b_n|^2b_n
-(\omega+i\Gamma)b_n
+h b_n^*,
\end{aligned}
\label{eq:autonomous_ab}
\end{equation}
for \(n\geq 1\), together with \(b_0=0\). Equation
\eqref{eq:autonomous_ab} defines a semi-infinite dimerized driven--damped
lattice, which is the natural setting for the analysis of edge-localized
states.
\subsection{Stationary edge states}
\label{subsec:stationary_edge_states}
We seek stationary solutions of the form
\begin{equation}
a_n(T)=A_n,
\qquad
b_n(T)=B_n,
\label{eq:stationary_ansatz}
\end{equation}
where \(\dot a_n=\dot b_n=0\). Substitution into
\eqref{eq:autonomous_ab} gives the stationary system
\begin{equation}
\begin{aligned}
0
&=
c_1 B_n+c_2 B_{n-1}
+|A_n|^2A_n
-(\omega+i\Gamma)A_n
+h A_n^*,\\[4pt]
0
&=
c_1 A_n+c_2 A_{n+1}
+|B_n|^2B_n
-(\omega+i\Gamma)B_n
+h B_n^*,
\end{aligned}
\label{eq:stationary_system}
\end{equation}
for \(n\geq 1\), with \(B_0=0\). We are interested in edge-localized
solutions, namely solutions \((A_n,B_n)\) satisfying
\begin{equation}
(A_n,B_n)\to (0,0)
\qquad
\text{as}
\qquad
n\to\infty.
\label{eq:edge_decay_condition}
\end{equation}
Let \((A_n,B_n)\) be a stationary solution of
\eqref{eq:stationary_system}. To study its spectral stability, we perturb it
as
\begin{equation}
\begin{aligned}
a_n(T)
&=
A_n+\delta\bigl(p_n e^{\lambda T}+q_n^*e^{\lambda^* T}\bigr),\\[4pt]
b_n(T)
&=
B_n+\delta\bigl(r_n e^{\lambda T}+s_n^*e^{\lambda^* T}\bigr),
\end{aligned}
\qquad
0<\delta\ll1,
\label{eq:perturbation_ansatz}
\end{equation}
where \(\lambda\in\mathbb{C}\) is the spectral parameter and
\((p_n,q_n,r_n,s_n)\) is the corresponding eigenvector.

Substituting \eqref{eq:perturbation_ansatz} into
\eqref{eq:autonomous_ab}, using the stationary equations
\eqref{eq:stationary_system}, and retaining terms of order
\(\mathcal{O}(\delta)\), we obtain
{\small \begin{equation}
\begin{aligned}
i\lambda p_n
&=
c_1 r_n+c_2 r_{n-1}
+\bigl(2|A_n|^2-\omega-i\Gamma\bigr)p_n
+\bigl(A_n^2+h\bigr)q_n,\\[4pt]
-i\lambda q_n
&=
c_1 s_n+c_2 s_{n-1}
+\bigl(2|A_n|^2-\omega+i\Gamma\bigr)q_n
+\bigl((A_n^*)^2+h\bigr)p_n,\\[4pt]
i\lambda r_n
&=
c_1 p_n+c_2 p_{n+1}
+\bigl(2|B_n|^2-\omega-i\Gamma\bigr)r_n
+\bigl(B_n^2+h\bigr)s_n,\\[4pt]
-i\lambda s_n
&=
c_1 q_n+c_2 q_{n+1}
+\bigl(2|B_n|^2-\omega+i\Gamma\bigr)s_n
+\bigl((B_n^*)^2+h\bigr)r_n,
\end{aligned}
\label{eq:linear_stability_problem}
\end{equation}}
for \(n\geq 1\), together with the boundary conditions
\begin{equation}
r_0=s_0=0.
\label{eq:stability_bc}
\end{equation}
Equivalently, the spectral problem may be written as
\begin{equation}
\lambda \mathbf{v}=\mathcal{L}\mathbf{v},
\qquad
\mathbf{v}^T=\left(p_n,q_n,r_n,s_n\right)_{n\geq 1},
\label{eq:matrix_evp}
\end{equation}
where \(\mathcal{L}\) denotes the linearization operator about the stationary
state.
For the driven--damped problem, a stationary edge state is spectrally
stable if all spectral points of the linearization lie strictly in the
left half-plane,
\[
\sup\{\operatorname{Re}(\lambda):\lambda\in\sigma(\mathcal{L})\}<0.
\]
In the conservative limit \(h=\Gamma=0\), this strict dissipative
criterion is replaced by the weaker spectral condition
\[
\sup\{\operatorname{Re}(\lambda):\lambda\in\sigma(\mathcal{L})\}\le 0.
\]

The spectrum of \(\mathcal{L}\) consists of isolated eigenvalues associated
with localized perturbations and the essential spectrum generated by the
linearization about the zero background. The latter provides the spectral
background against which edge eigenvalues and their stability must be
interpreted.
\subsection{Linear spectrum about the zero solution}
\label{subsec:linear_zero}

Before analyzing nonlinear stationary branches, we ex-
amine the spectrum of the linearized problem about the
trivial state. Setting \(A_n=B_n=0\) in \eqref{eq:linear_stability_problem} gives
\begin{equation}
\begin{aligned}
i\lambda p_n
&=
c_1 r_n+c_2 r_{n-1}
-(\omega+i\Gamma)p_n
+h q_n,\\[4pt]
-i\lambda q_n
&=
c_1 s_n+c_2 s_{n-1}
-(\omega-i\Gamma)q_n
+h p_n,\\[4pt]
i\lambda r_n
&=
c_1 p_n+c_2 p_{n+1}
-(\omega+i\Gamma)r_n
+h s_n,\\[4pt]
-i\lambda s_n
&=
c_1 q_n+c_2 q_{n+1}
-(\omega-i\Gamma)s_n
+h r_n,
\end{aligned}
\label{eq:zero_linear_system}
\end{equation}
for \(n\geq 1\).

\subsubsection*{Essential spectrum}
To determine the essential spectrum, we use the Bloch ansatz
\begin{equation}
\begin{pmatrix}
p_n\\
q_n\\
r_n\\
s_n
\end{pmatrix}
=
e^{ikn}
\begin{pmatrix}
P\\
Q\\
R\\
S
\end{pmatrix},
\qquad
k\in[-\pi,\pi].
\label{eq:bloch_ansatz}
\end{equation}
Substitution into \eqref{eq:zero_linear_system} yields
\begin{equation}
\begin{aligned}
i\lambda P
&=
\overline{C(k)}\,R-(\omega+i\Gamma)P+hQ,\\[4pt]
-i\lambda Q
&=
\overline{C(k)}\,S-(\omega-i\Gamma)Q+hP,\\[4pt]
i\lambda R
&=
C(k)\,P-(\omega+i\Gamma)R+hS,\\[4pt]
-i\lambda S
&=
C(k)\,Q-(\omega-i\Gamma)S+hR,
\end{aligned}
\label{eq:bloch_algebraic}
\end{equation}
where
\begin{equation}
C(k)=c_1+c_2e^{ik},
\qquad
k\in[-\pi,\pi].
\label{eq:Ck_def}
\end{equation}
Since the SSH coupling block has eigenvalues \(\pm |C(k)|\), the dispersion
relation consists of the four branches
\begin{equation}
\lambda_{\sigma,\eta}(k)
=
-\Gamma
+\sigma\sqrt{
h^2-\bigl(\omega-\eta |C(k)|\bigr)^2
},
\label{eq:bulk_dispersion}
\end{equation}
where $\sigma,\eta\in\{+1,-1\}.$
Consequently, the essential spectrum is
\begin{equation}
\sigma_{\mathrm{ess}}(\mathcal{L})
=
\bigcup_{\eta=\pm 1}
\left\{
-\Gamma
\pm
\sqrt{
h^2-\bigl(\omega-\eta |C(k)|\bigr)^2
}
:
k\in[-\pi,\pi]
\right\}.
\label{eq:essential_spectrum}
\end{equation}
If
\begin{equation}
|h|
<
\min_{\eta=\pm1}
\min_{k\in[-\pi,\pi]}
\left|
\omega-\eta |C(k)|
\right|,
\label{eq:h_gap_condition}
\end{equation}
then the square root in \eqref{eq:bulk_dispersion} is purely imaginary for
all \(k\). In this case, the essential spectrum lies on the vertical line
\begin{equation}
\operatorname{Re}(\lambda)=-\Gamma.
\label{eq:bulk_vertical}
\end{equation}
\begin{figure}[tbhp!]
\centering
\includegraphics[scale=0.5]{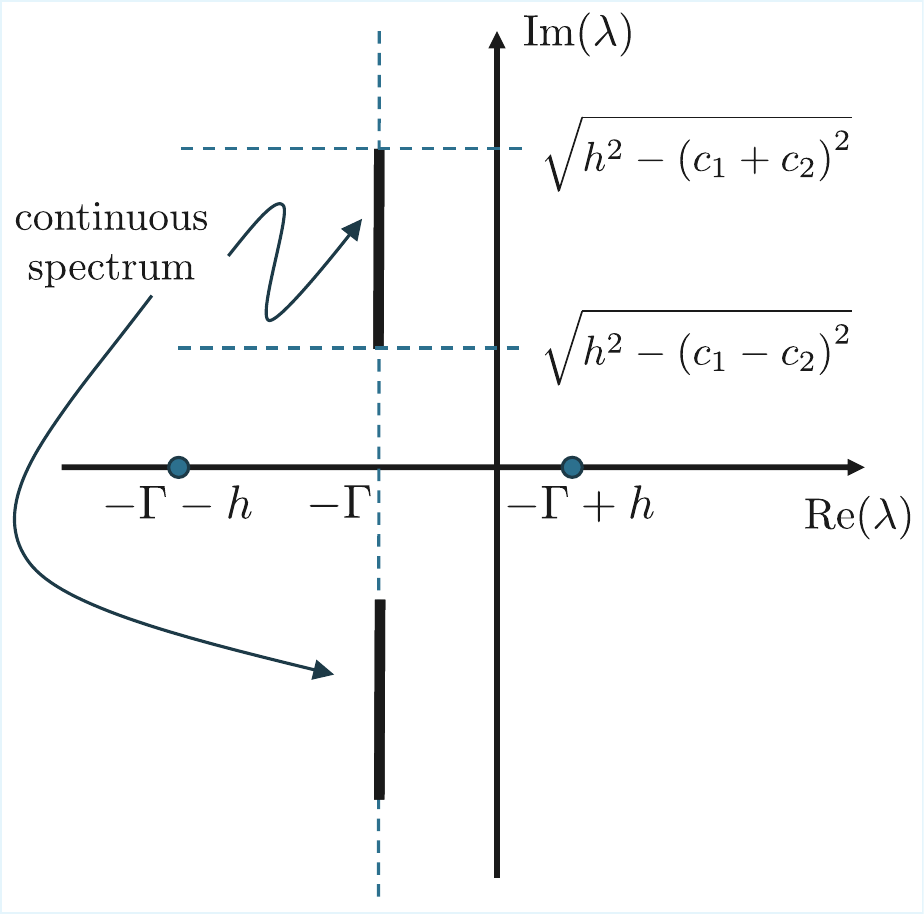}
\caption{Schematic spectrum of the linearized operator about the zero
solution when \(\omega=0\). The essential spectrum is determined by
\eqref{eq:bulk_dispersion}. In the regime \eqref{eq:h_gap_condition}, it lies
on the vertical line \(\operatorname{Re}(\lambda)=-\Gamma\).}
\label{fig:spectrum_zero}
\end{figure}
\subsubsection*{Edge spectrum}
For the semi-infinite lattice, the left boundary allows for exponentially localized edge modes. We consider an ansatz supported on the \(A\)-sublattice,
	\begin{equation}
		\begin{aligned}
			& r_n=s_n=0,\\[4pt]
			& p_n=P\rho^{\,n-1},
			\qquad
			q_n=Q\rho^{\,n-1},
			\qquad
			|\rho|<1.
		\end{aligned}
		\label{eq:left_edge_linear_mode}
	\end{equation}
Substituting \eqref{eq:left_edge_linear_mode} into
\eqref{eq:zero_linear_system} gives
\begin{equation}
\begin{aligned}
\bigl(\omega+i(\lambda+\Gamma)\bigr)P-hQ&=0,\\[4pt]
-hP+\bigl(\omega-i(\lambda+\Gamma)\bigr)Q&=0,\\[4pt]
P(c_1+c_2\rho)&=0,\\[4pt]
Q(c_1+c_2\rho)&=0.
\end{aligned}
\label{eq:edge_algebraic_system}
\end{equation}
For a nontrivial vector \((P,Q)^T\neq(0,0)\), the last two equations in
\eqref{eq:edge_algebraic_system} imply
\begin{equation}
c_1+c_2\rho=0,
\qquad
\text{hence}
\qquad
\rho=-\frac{c_1}{c_2}.
\label{eq:rho_linear_edge}
\end{equation}
The localization condition \(|\rho|<1\) therefore becomes
\begin{equation}
\left|\frac{c_1}{c_2}\right|<1.
\label{eq:topological_condition_linear}
\end{equation}
For positive couplings \(c_1,c_2>0\), this is equivalent to
\begin{equation}
c_1<c_2.
\label{eq:topological_condition_positive}
\end{equation}
The first two equations in \eqref{eq:edge_algebraic_system} may be written as
\begin{equation}
\begin{pmatrix}
\omega+i(\lambda+\Gamma) & -h\\[4pt]
-h & \omega-i(\lambda+\Gamma)
\end{pmatrix}
\begin{pmatrix}
P\\[2pt]
Q
\end{pmatrix}
=
\begin{pmatrix}
0\\[2pt]
0
\end{pmatrix}.
\label{eq:edge_matrix_system}
\end{equation}
A nontrivial solution exists if
\begin{equation}
\det
\begin{pmatrix}
\omega+i(\lambda+\Gamma) & -h\\[4pt]
-h & \omega-i(\lambda+\Gamma)
\end{pmatrix}
=0.
\label{eq:edge_determinant_condition}
\end{equation}
This gives
\begin{equation}
\omega^2+(\lambda+\Gamma)^2-h^2=0.
\label{eq:edge_characteristic_equation}
\end{equation}
Thus, the edge eigenvalues are
\begin{equation}
\lambda_{\mathrm{edge},\pm}
=
-\Gamma\pm\sqrt{h^2-\omega^2}.
\label{eq:edge_eigs_zero}
\end{equation}
We conclude that the linear spectrum about the zero solution consists of the
essential spectrum \eqref{eq:essential_spectrum} and, in the topological
regime \eqref{eq:topological_condition_linear}, the discrete edge eigenvalues
\eqref{eq:edge_eigs_zero}.
\subsection{Onset of stationary edge states}
\label{subsec:onset}
We now determine where a nonlinear stationary edge branch bifurcates from the
zero solution. Since
\begin{equation}
A_n=B_n=0
\label{eq:onset_zero_solution}
\end{equation}
is always a solution of the stationary system \eqref{eq:stationary_system},
the onset of a nontrivial branch is determined by the loss of invertibility
of the linearized stationary problem.
Linearizing \eqref{eq:stationary_system} about \(A_n=B_n=0\) gives
\begin{equation}
\begin{aligned}
0
&=
c_1 B_n+c_2 B_{n-1}
-(\omega+i\Gamma)A_n
+hA_n^*,\\[4pt]
0
&=
c_1 A_n+c_2 A_{n+1}
-(\omega+i\Gamma)B_n
+hB_n^*.
\end{aligned}
\label{eq:linearized_stationary}
\end{equation}
	Motivated by the linear edge modes identified in Sec.~\ref{subsec:linear_zero}, we consider solutions localized at the left boundary and supported on the \(A\)-sublattice,
	\begin{equation}
		B_n=0,\qquad A_n=A\rho^{\,n-1},\qquad |\rho|<1.
		\label{eq:left_edge_stationary_ansatz}
	\end{equation}
	As shown previously, substitution into the second equation of \eqref{eq:linearized_stationary} yields \eqref{eq:rho_linear_edge} and \eqref{eq:topological_condition_linear}.
	
	Substituting \eqref{eq:left_edge_stationary_ansatz} into the first equation of \eqref{eq:linearized_stationary} gives
	\begin{equation}
		-(\omega+i\Gamma)A+hA^*=0.
		\label{eq:linear_amplitude}
	\end{equation}
Writing
\begin{equation}
A=\alpha+i\beta,
\qquad
A^*=\alpha-i\beta,
\qquad
\alpha,\beta\in\mathbb{R},
\label{eq:onset_complex_decomposition}
\end{equation}
and separating the real and imaginary parts yields 
\begin{equation}
\begin{aligned}
(h-\omega)\alpha+\Gamma\beta&=0,\\[4pt]
-\Gamma\alpha-(h+\omega)\beta&=0.
\end{aligned}
\label{eq:onset_real_system}
\end{equation}
A nontrivial solution \((\alpha,\beta)\neq(0,0)\) exists precisely when
\begin{equation}
\det
\begin{pmatrix}
h-\omega & \Gamma\\[4pt]
-\Gamma & -(h+\omega)
\end{pmatrix}
=0.
\label{eq:onset_determinant}
\end{equation}
Therefore,
\begin{equation}
-(h-\omega)(h+\omega)+\Gamma^2=0,
\label{eq:onset_determinant_expanded}
\end{equation}
or equivalently
\begin{equation}
\omega^2=h^2-\Gamma^2.
\label{eq:onset_threshold}
\end{equation}
A real onset point exists only when $|h| > \Gamma$.  In that case, a nonlinear stationary edge branch bifurcates from the zero solution
when
\begin{equation}
\omega^2=h^2-\Gamma^2
\label{eq:onset_threshold_final}
\end{equation}
and
\begin{equation}
\left|\frac{c_1}{c_2}\right|<1
\label{eq:onset_localization_final}
\end{equation}
hold simultaneously. For \(c_1,c_2>0\), the localization condition
\eqref{eq:onset_localization_final} reduces to \(c_1<c_2\).
The same onset condition follows from the linear edge eigenvalues
\begin{equation}
\lambda_{\mathrm{edge},\pm}
=
-\Gamma\pm\sqrt{h^2-\omega^2}.
\label{eq:onset_edge_eigenvalues_recalled}
\end{equation}
Indeed, the branch emerges when an edge eigenvalue reaches the origin,
\begin{equation}
\lambda_{\mathrm{edge},+}=0,
\label{eq:onset_zero_eigenvalue_condition}
\end{equation}
which is equivalent to
\begin{equation}
\omega^2=h^2-\Gamma^2.
\label{eq:onset_threshold_equivalence}
\end{equation}
Hence, the onset of nonlinear stationary edge states is associated with a
marginal stability condition of the corresponding linear edge mode.
In the next section, we continue these stationary edge branches into the
nonlinear regime and examine their spectral stability.

\section{Approximate Edge Soliton Solutions}
\label{sec:edge_solitons}
	
In this section, we derive an analytical approximation for the localized edge solutions of the driven--damped SSH amplitude system. The construction is inspired by the approximation developed by Ma and Susanto for nonlinear
SSH edge solitons \cite{ma2021topological}. The main idea is to approximate the edge state by an exponentially decaying spatial profile and to assume that
the nonlinear contribution is dominant only at the first edge site, while the remaining sites are treated linearly.
	
We seek a left-localized solution of \eqref{eq:stationary_system} on the semi-infinite lattice, subject to the left boundary conditions
	\[
	B_0=0,\qquad (A_n,B_n) \to (0,0) \,\, \text{as} \,\,n \to \infty.
	\]
	We assume that the solution is in the form
	\begin{align}
		A_n &\approx \widetilde{A}(-L)^{n-1}, &
		B_n &\approx \widetilde{B}(-L)^n,
		\label{eq:localized_stat_sol}
	\end{align}
	where \(0<L<1\) is the spatial decay factor. Since the driven--damped system
	is complex-valued, we write
	\begin{align}
		\widetilde{A}
		&=
		\widetilde{\alpha}
		+i\widetilde{\beta},
		&
		\widetilde{B}
		&=
		\widetilde{\eta}
		+i\widetilde{\xi}.
		\label{eq:Atilde_Btilde_real}
	\end{align}
	
	The approximation is obtained under the assumption that the only nonlinear
	equation retained at leading order is the equation at the first edge site
	\(A_1\). Thus, for \(n\geq 2\), the equation for \(A_n\) is linearized, and
	the equations for \(B_n\) are also kept linear. Substituting
	\eqref{eq:localized_stat_sol} into the linear part of
	\eqref{eq:stationary_system} for \(n\geq 2\) gives
	\begin{align}
		(\omega+i\Gamma)\widetilde{A}
		-h\widetilde{A}^*
		=
		(c_2-c_1L)\widetilde{B}, \label{eq:linear_A_complex}\\
		(\omega+i\Gamma)\widetilde{B}
		-h\widetilde{B}^*
		=
		\left(c_2-\frac{c_1}{L}\right)\widetilde{A}. \label{eq:linear_B_complex}
	\end{align}
	
	Using \eqref{eq:Atilde_Btilde_real}, equations
	\eqref{eq:linear_A_complex}--\eqref{eq:linear_B_complex} can be written as
	a real four-dimensional eigenvalue problem
	\begin{align}
		\omega\Psi
		=
		\mathcal{M}(L)\Psi,
		\label{eq:four_dim_eigenproblem}
	\end{align}
	where
	\[
	\Psi
	=
	\begin{pmatrix}
		\widetilde{\alpha}\\
		\widetilde{\beta}\\
		\widetilde{\eta}\\
		\widetilde{\xi}
	\end{pmatrix},
	\]
	and
	\begin{align}
		\mathcal{M}(L)
		=
		\begin{pmatrix}
			h & \Gamma & c_2-c_1L & 0\\
			-\Gamma & -h & 0 & c_2-c_1L\\
			c_2-\dfrac{c_1}{L} & 0 & h & \Gamma\\
			0 & c_2-\dfrac{c_1}{L} & -\Gamma & -h
		\end{pmatrix}.
		\label{eq:Mtilde_L}
	\end{align}
	
	For compactness, define
	\begin{align}
		p(L)
		&=
		c_2-c_1L,
		&
		q(L)
		&=
		c_2-\frac{c_1}{L},
		&
		g(L)
		&=
		p(L)q(L).
		\label{eq:pqg_def}
	\end{align}
	Then the matrix \(\mathcal{M}(L)\) has the block form
	\begin{align}
		\mathcal{M}(L)
		=
		\begin{pmatrix}
			J & p(L)I_2\\
			q(L)I_2 & J
		\end{pmatrix},
		\qquad
		J=
		\begin{pmatrix}
			h & \Gamma\\
			-\Gamma & -h
		\end{pmatrix}.
		\label{eq:block_matrix}
	\end{align}
	Since the off-diagonal blocks are scalar multiples of the identity, the
	characteristic determinant reduces to
	\begin{align}
		\det\left(\omega I_4-\mathcal{M}(L)\right)
		&=
		\det\left[
		\left(\omega I_2-J\right)^2
		-g(L)I_2
		\right].
		\label{eq:det_reduction}
	\end{align}
	The eigenvalues of \(J\) are
	\[
	\pm \chi,
	\qquad
	\chi=\sqrt{h^2-\Gamma^2}.
	\]
	Therefore, provided \(|h|>\Gamma\), the characteristic determinant factors as
	\begin{align}
		\det\left(\omega I_4-\mathcal{M}(L)\right)
		=
		\left[
		(\omega-\chi)^2-g(L)
		\right]
		\left[
		(\omega+\chi)^2-g(L)
		\right].
		\label{eq:det_factorized}
	\end{align}
	Thus, the two phase-locked branches are determined by
	\begin{align}
		g(L)
		=
		\left(\omega-s\chi\right)^2,
		\qquad
		s=\pm1.
		\label{eq:gL_branch}
	\end{align}
	
	Let
	\begin{align}
		\nu_s
		=
		\omega-s\sqrt{h^2-\Gamma^2},
		\qquad
		s=\pm1.
		\label{eq:nu_s_def}
	\end{align}
	Then the localization factor satisfies
	\begin{align}
		\nu_s^2
		=
		\left(c_2-c_1L\right)
		\left(c_2-\frac{c_1}{L}\right).
		\label{eq:L_relation}
	\end{align}
	Equivalently,
	\begin{align}
		L^2
		-
		\frac{c_1^2+c_2^2-\nu_s^2}{c_1c_2}L
		+1
		=
		0.
		\label{eq:L_quadratic}
	\end{align}
	Hence,
	\begin{align}
		L_s
		=
		\frac{
			c_1^2+c_2^2-\nu_s^2
			-
			\sqrt{
				\nu_s^4
				-2\nu_s^2(c_1^2+c_2^2)
				+
				(c_1^2-c_2^2)^2
			}
		}
		{2c_1c_2}.
		\label{eq:Ls_final}
	\end{align}
	The sign is chosen such that \(0<L_s<1\). In the topological regime
	\(0<c_1<c_2\), this branch satisfies
	\[
	L_s\to \frac{c_1}{c_2}
	\qquad
	\text{as}
	\qquad
	\nu_s\to0,
	\]
	which recovers the decay factor of the linear SSH edge mode.
	
	The localization condition may also be written as
	\begin{align}
		|\nu_s|
		<
		|c_2-c_1|.
		\label{eq:localization_condition}
	\end{align}
	Thus, in the driven--damped system, the effective spectral distance that
	controls the localization is not simply \(\omega\), but rather
	\(\nu_s=\omega-s\sqrt{h^2-\Gamma^2}\).
	
	We next determine the approximate amplitudes \(\widetilde{A}\) and
	\(\widetilde{B}\). The eigenvector of the onsite block \(J\) corresponding
	to the eigenvalue \(s\chi\) is
	\[
	\begin{pmatrix}
		1\\
		\tau_s
	\end{pmatrix},
	\qquad
	\tau_s
	=
	\frac{-h+s\sqrt{h^2-\Gamma^2}}{\Gamma}.
	\label{eq:tau_s}
	\]
	Accordingly, the associated normalized complex phase is
	\begin{align}
		\zeta_s
		=
		\frac{1+i\tau_s}{\sqrt{1+\tau_s^2}}.
		\label{eq:zeta_s}
	\end{align}
	
	On the branch \(s=\pm1\), the first two components of
	\eqref{eq:four_dim_eigenproblem} yield
	\begin{align}
		\nu_s\widetilde{A}
		=
		p(L_s)\widetilde{B},
		\label{eq:A_B_relation_1}
	\end{align}
	whereas the last two components yield
	\begin{align}
		\nu_s\widetilde{B}
		=
		q(L_s)\widetilde{A}.
		\label{eq:A_B_relation_2}
	\end{align}
	Therefore,
	\begin{align}
		\widetilde{B}
		=
		\frac{q(L_s)}{\nu_s}\widetilde{A}
		=
		\frac{c_2-\dfrac{c_1}{L_s}}{\nu_s}\widetilde{A}.
		\label{eq:B_in_terms_A}
	\end{align}
	
	It remains to determine \(\widetilde{A}\) from the nonlinear edge equation.
	At \(n=1\), using \(B_0=0\) and \(B_1=-L_s\widetilde{B}\), the first equation of
	\eqref{eq:stationary_system} gives
	\begin{align}
		\omega\widetilde{A}
		=
		-c_1L_s\widetilde{B}
		+
		|\widetilde{A}|^2\widetilde{A}
		-i\Gamma\widetilde{A}
		+h\widetilde{A}^*.
		\label{eq:edge_A1_full}
	\end{align}
	On the phase-locked branch, the onsite linear part satisfies
	\[
	-i\Gamma\widetilde{A}
	+h\widetilde{A}^*
	=
	s\chi\widetilde{A}.
	\]
	Hence \eqref{eq:edge_A1_full} can be written as
	\begin{align}
		\nu_s\widetilde{A}
		=
		-c_1L_s\widetilde{B}
		+
		|\widetilde{A}|^2\widetilde{A}.
		\label{eq:edge_A1_branch}
	\end{align}
	In the leading Ma--Susanto approximation, the edge-site balance is assumed
	to be dominated by the onsite cubic term at \(A_1\), and the coupling
	correction \(-c_1L_s\widetilde{B}\) is neglected at leading order. Thus,
	\begin{align}
		\nu_s\widetilde{A}
		\approx
		|\widetilde{A}|^2\widetilde{A}.
		\label{eq:leading_edge_balance}
	\end{align}
	For a nontrivial solution, this gives
	\begin{align}
		|\widetilde{A}|^2
		\approx
		\nu_s.
		\label{eq:Atilde_modulus}
	\end{align}
	Therefore, for an admissible branch satisfying \(\nu_s>0\), we obtain
	\begin{align}
		\widetilde{A}
		&\approx
		\sqrt{\nu_s}\,\zeta_s,
		\label{eq:Atilde_final}\\
		\widetilde{B}
		&\approx
		\frac{c_2-\dfrac{c_1}{L_s}}{\sqrt{\nu_s}}\,
		\zeta_s.
		\label{eq:Btilde_final}
	\end{align}
	Equivalently, in terms of the real components,
	\begin{align}
		\widetilde{\alpha}
		&=
		\frac{\sqrt{\nu_s}}{\sqrt{1+\tau_s^2}},
		&
		\widetilde{\beta}
		&=
		\frac{\tau_s\sqrt{\nu_s}}{\sqrt{1+\tau_s^2}},
		\label{eq:alpha_beta_final}\\
		\widetilde{\eta}
		&=
		\frac{c_2-\dfrac{c_1}{L_s}}{\sqrt{\nu_s}}
		\frac{1}{\sqrt{1+\tau_s^2}},
		&
		\widetilde{\xi}
		&=
		\frac{c_2-\dfrac{c_1}{L_s}}{\sqrt{\nu_s}}
		\frac{\tau_s}{\sqrt{1+\tau_s^2}}.
		\label{eq:eta_xi_final}
	\end{align}
	
Consequently, the approximate left-localized edge soliton is given by
\begin{align}
	A_n
	&\approx
	\sqrt{\nu_s}\,\zeta_s(-L_s)^{n-1},
	\label{eq:An_edge_final}\\
	B_n
	&\approx
	\frac{c_2-\dfrac{c_1}{L_s}}{\sqrt{\nu_s}}\,
	\zeta_s(-L_s)^n.
	\label{eq:Bn_edge_final}
\end{align}
The approximation is valid when \(0<L_s<1\), \(\nu_s>0\), and the
localization condition \eqref{eq:localization_condition} holds. 

An important consequence of the driven--damped setting is the emergence of
two distinct phase-locked edge-soliton families corresponding to the
two choices \(s=\pm1\). These families are characterized by the
effective detuning parameters $\nu_s$ \eqref{eq:nu_s_def}, which replace the single detuning parameter \(\omega\) of the conservative SSH lattice. In the limit \(h=\Gamma=0\), the two branches coalesce and recover the conservative edge-soliton family. For
nonvanishing drive and damping, however, the two phase-locked branches
remain distinct and correspond to different phase relations with the
external drive. As will be shown in Sec.~\ref{sec:numerics}, these two dissipative edge
states possess markedly different stability properties: one branch is
typically unstable throughout most of its existence region, whereas the
other develops substantially larger stability intervals and remains
dynamically robust even when weakly unstable.

Finally, the corresponding approximate total power is
\begin{align}
	P
	&=
	\sum_{n\geq1}
	\left(
	|A_n|^2+|B_n|^2
	\right) \notag\\
	&\approx
	\sum_{n\geq1}
	\left(
	|\widetilde{A}|^2L_s^{2(n-1)}
	+
	|\widetilde{B}|^2L_s^{2n}
	\right) \notag\\
	&=
	\frac{
		|\widetilde{A}|^2
		+
		|\widetilde{B}|^2L_s^2
	}
	{1-L_s^2}.
	\label{eq:power_edge_approx}
\end{align}
Equation~\eqref{eq:power_edge_approx} provides an explicit analytical
approximation for the power of the nonlinear edge branch, and will be
used in Sec.~\ref{sec:numerics} to compare with the numerically continued solutions. The
dependence of \(P\) on the phase-locking parameter \(s\) enters through
\(L_s\) and \(\nu_s\), allowing the two dissipative edge-soliton
families to be distinguished quantitatively already at the level of the
analytical approximation.

\section{Numerical Results}
\label{sec:numerics}

In this section, we present numerical computations of stationary edge states, their spectral stability, and their time-dependent dynamics. The purpose is threefold: first, to verify the analytical approximation derived in Sec.~\ref{sec:edge_solitons}; second, to continue the nonlinear edge branch away from onset; and third, to determine which portions of the branch are spectrally and dynamically relevant.
Since the computations are performed on a finite lattice, we truncate the semi-infinite problem to a finite computational domain. We consider \(n=1,\ldots,N\) and impose the boundary conditions
\begin{equation}
	b_0=b_N=0.
	\label{eq:numerical_bc}
\end{equation}
This finite-dimensional truncation provides an accurate approximation of the semi-infinite problem whenever the stationary edge state is sufficiently localized near the left boundary.
Stationary solutions are computed by solving the nonlinear algebraic system \eqref{eq:stationary_system} subject to \eqref{eq:numerical_bc}. The computation is carried out using a Newton-type method implemented through the MATLAB routine \texttt{fsolve}. In practice, the complex variables are decomposed into their real and imaginary parts, so that the stationary problem is written as a real nonlinear system. Numerical continuation in a control parameter, such as \(\omega\), is performed by using a previously computed stationary solution as the initial guess for the next parameter value.
Once a stationary state is obtained, its spectral stability is determined from the corresponding linear eigenvalue problem \eqref{eq:linear_stability_problem}. On the truncated domain, this yields a finite-dimensional non-Hermitian matrix eigenvalue problem. A stationary solution is classified as spectrally stable if the real parts of all eigenvalues are nonpositive up to numerical tolerance, while the presence of an eigenvalue with a positive real part indicates spectral instability.
For selected unstable or weakly stable solutions, we also examine the time-dependent dynamics by direct numerical integration of the SSH amplitude system \eqref{eq:autonomous_ab}. The same finite domain and boundary conditions are used. Time integration is carried out using the classical fourth-order Runge--Kutta method. The initial condition is taken as the stationary solution perturbed by a small-amplitude disturbance, either along a relevant eigenmode or by small random noise. This allows us to compare the spectral prediction with the observed nonlinear evolution.

\subsection{Power versus detuning}
\label{subsec:power_vs_omega}

We first examine how the stationary edge-state branches depend on the detuning parameter \(\omega\). To quantify the branches, we monitor the total power $P$ \eqref{eq:power_edge_approx}. For the driven-damped problem, the onset condition obtained from the analytical calculation is $\nu_s=0$, i.e., 
\[
	h^2=\omega^2+\Gamma^2,
\]
which gives
\begin{equation}
	\omega_c=\pm\sqrt{h^2-\Gamma^2}.
	\label{eq:omega_c_numerics}
\end{equation}
In the conservative case \(h=\Gamma=0\), the corresponding onset occurs at \(\omega_c=0\).

\begin{figure*}[tbhp]
	\centering
	\subfloat[$\Gamma =0$, $h =0$, $c_1=0.6$]{\includegraphics[scale=0.4]{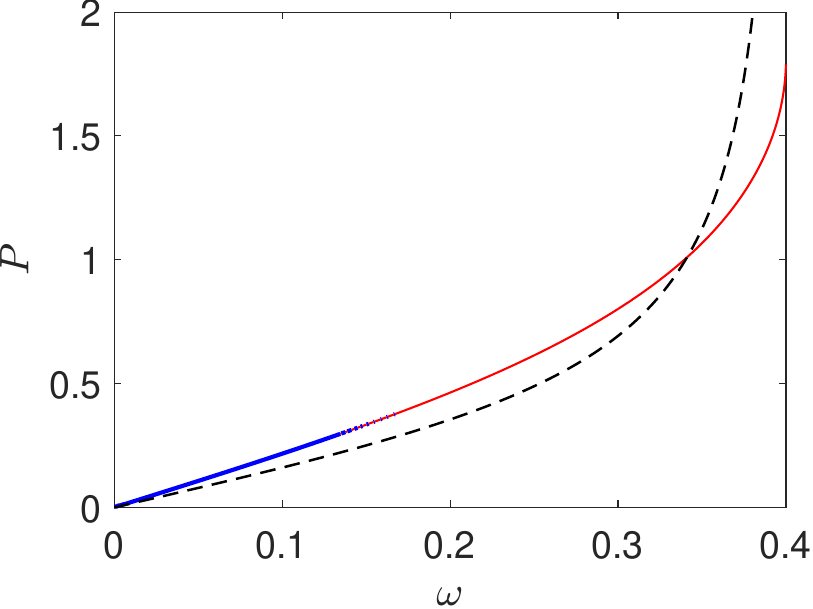}\label{bifur_1}}\,\,
	\subfloat[$\Gamma =0.04$, $h =0.05$, $c_1=0.6$]{\includegraphics[scale=0.4]{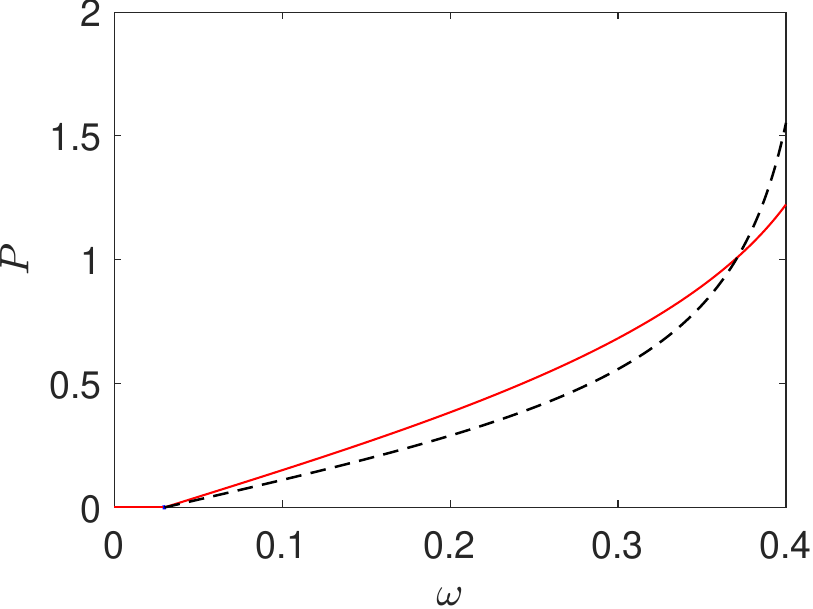}\label{bifur_2}}\,\,
	\subfloat[$\Gamma =0.04$, $h =-0.05$, $c_1=0.6$]{\includegraphics[scale=0.4]{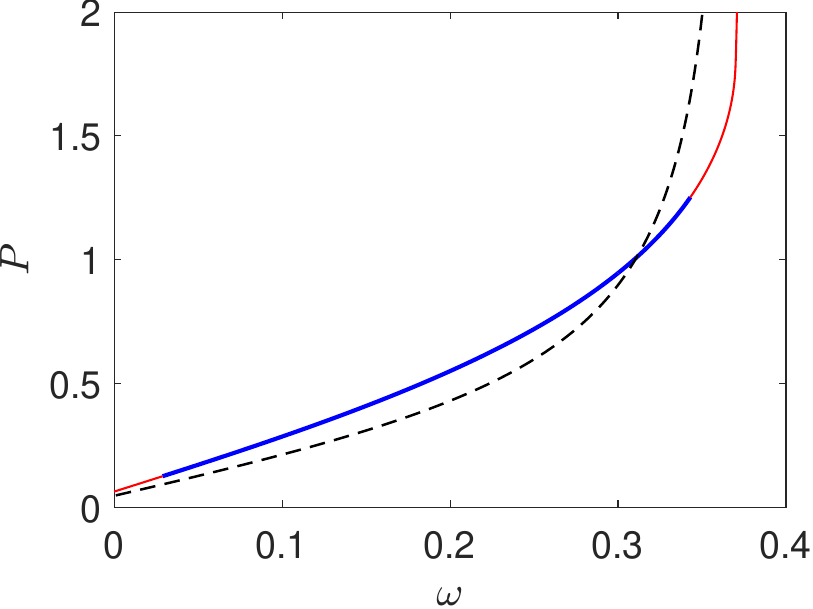}\label{bifur_3}}
	\caption{Bifurcation diagram showing the total power \(P\) as a function of the detuning parameter \(\omega\). The solid curve denotes the numerical continuation, while the dashed curve denotes the analytical approximation. The panels correspond to \(c_2=1\). For the driven--damped cases in panels (b) and (c), the predicted onset is \(|\omega_c|=\sqrt{h^2-\Gamma^2}=0.03\); in the conservative case, the corresponding onset occurs at \(\omega_c=0\). The blue and red portions correspond to stable and unstable solutions, respectively.}
	\label{fig:power_vs_omega}
\end{figure*}

Figure~\ref{fig:power_vs_omega} shows the total power \(P\) as a function of \(\omega\) for \(h=\pm0.05\) and \(\Gamma=0.04\). From \eqref{eq:omega_c_numerics}, the predicted onset satisfies $|\omega_c|=	\sqrt{0.05^2-0.04^2}=0.03$. For comparison, we also consider the case in which the system is undamped and undriven. The numerical continuation is compared with the analytical approximation derived in Sec.~\ref{sec:edge_solitons}. Near onset, the two curves agree very well, as expected from the small-amplitude construction. As the branch extends farther from the threshold, the power increases, and a visible discrepancy emerges between the two curves. This deviation is natural because the analytical approximation retains only the leading-order localized contribution and neglects higher-order corrections to the bulk tail. The monotone growth of \(P\) along the branch indicates that the edge state becomes progressively more energetic as it is continued away from the onset. 

\subsection{Stationary edge profiles and spectral stability}
\label{subsec:profiles_stability}

We next examine representative stationary edge states selected from the nonlinear branches shown in Fig.~\ref{fig:power_vs_omega}. Specifically, we choose solutions with
\[
\omega=0.37,
\qquad
c_1=0.6,
\qquad
c_2=1,
\]
for the conservative system \((\Gamma,h)=(0,0)\) and for the two driven--damped systems \((\Gamma,h)=(0.04,\pm0.05)\). We also examine the spectral stability of the stationary edge states. For each numerically computed stationary solution, we solve the eigenvalue problem \eqref{eq:linear_stability_problem}.

The selected solutions all belong to spectrally unstable portions of the corresponding branches. They are chosen to illustrate the different instability mechanisms that arise in the conservative and driven--damped lattices and to provide representative examples for the dynamical simulations presented below.

Figure~\ref{fig:prof} displays the real and imaginary parts of the stationary edge-state profiles together with the spectra of the associated linearization operator \eqref{eq:matrix_evp}. In all cases, the solutions remain strongly localized near the left boundary, confirming their interpretation as nonlinear continuations of the topological edge mode. The spectra reveal the nature of the instability by showing the locations of the eigenvalues in the complex plane.

The instability mechanism depends strongly on the parameter regime. Panels~(a) and~(c) exhibit oscillatory instabilities associated with complex-conjugate eigenvalue pairs having positive real parts. In contrast, panel~(b) exhibits an exponential instability caused by a real positive eigenvalue. Despite these differences, all three solutions maintain a pronounced edge-localized structure.

The comparison between panels~(b) and~(c) is particularly noteworthy. Although both solutions are spectrally unstable, the unstable eigenvalues in panel~(c) lie much closer to the imaginary axis, indicating a substantially weaker instability. As will be shown in Sec.~\ref{subsec:time_dynamics}, this weak instability has little effect on the observable dynamics over long time intervals, leading to a remarkably robust edge-localized state. This observation already suggests that the combined action of parametric driving and damping can significantly enhance the robustness of nonlinear topological edge solitons.

\begin{figure*}[tbhp]
	\centering
	\subfloat[$\Gamma =0$, $h =0$, $c_1=0.6$]{\includegraphics[scale=0.4]{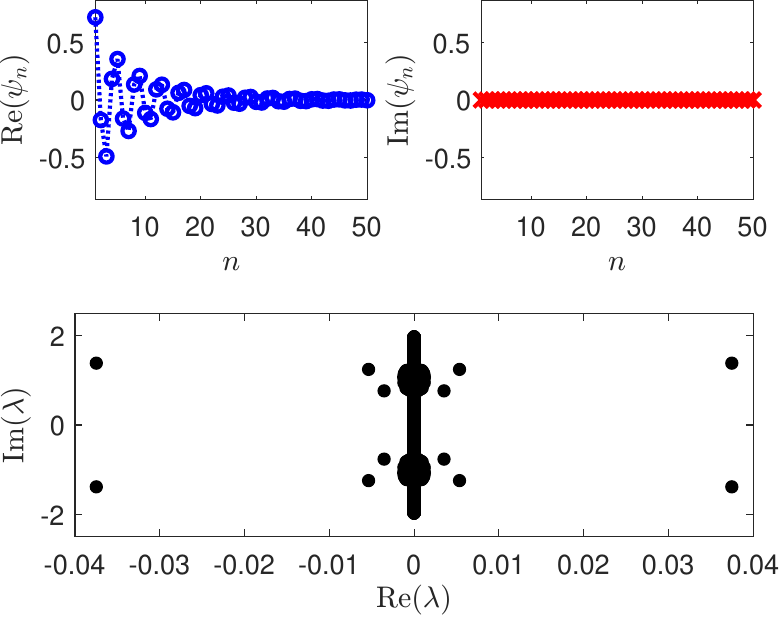}\label{profile_psi_1}}\,\,
	\subfloat[$\Gamma =0.04$, $h =0.05$, $c_1=0.6$]{\includegraphics[scale=0.4]{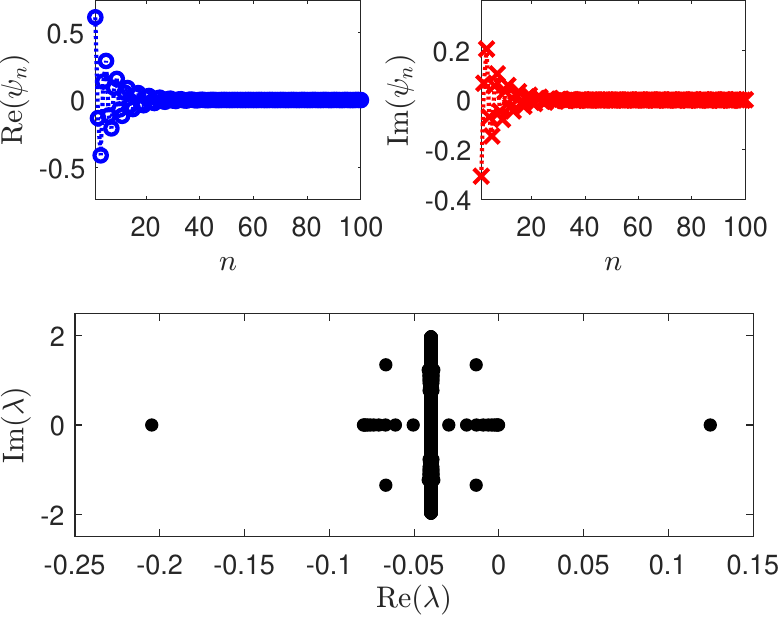}\label{profile_psi_2}}\,\,
	\subfloat[$\Gamma =0.04$, $h =-0.05$, $c_1=0.6$]{\includegraphics[scale=0.4]{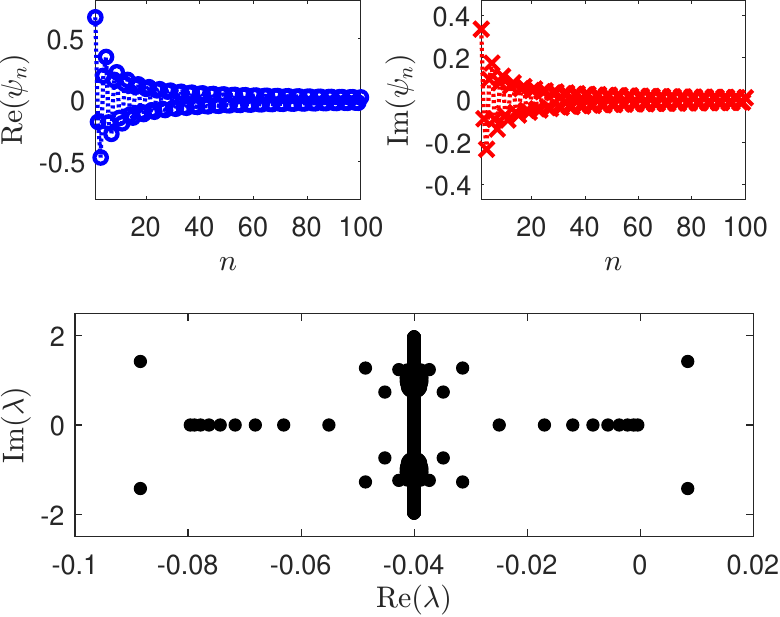}\label{profile_psi_3}}
	\caption{Representative stationary edge states and their linearization spectra for \(\omega=0.37\) and \(c_2=1\). Each group of panels shows \(\operatorname{Re}(A_n)\), \(\operatorname{Im}(A_n)\), and the corresponding spectrum of the linearized operator. The parameters are (a) \(\Gamma=0\), \(h=0\), \(c_1=0.6\), (b) \(\Gamma=0.04\), \(h=0.05\), \(c_1=0.6\), and (c) \(\Gamma=0.04\), \(h=-0.05\), \(c_1=0.6\). The selected solutions are spectrally unstable. Panels (a) and (c) exhibit oscillatory instabilities associated with complex unstable eigenvalues, whereas panel (b) exhibits an exponential instability associated with a real positive eigenvalue.}
	\label{fig:prof}
\end{figure*}

\begin{figure*}[tbhp]
	\centering
	\subfloat[$\Gamma =0$, $h =0$]{\includegraphics[scale=0.4]{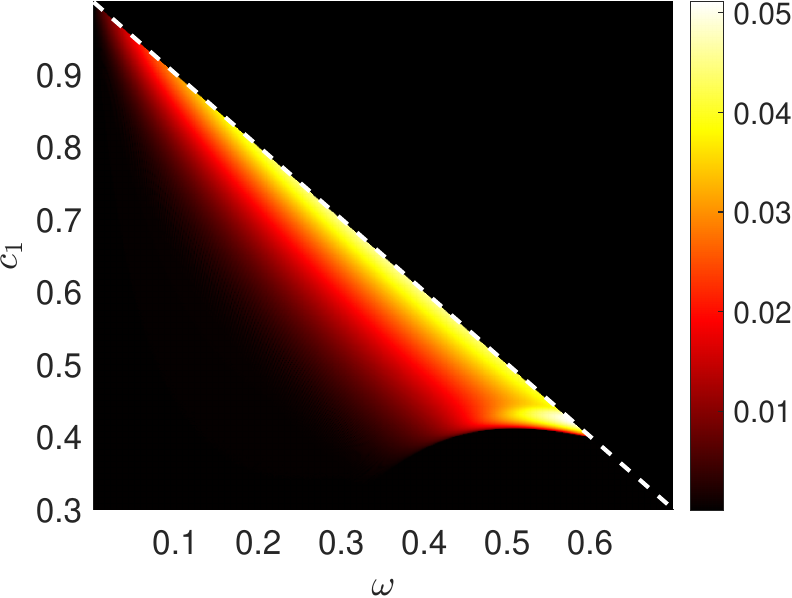}\label{heatmap1}}\,\,
	\subfloat[$\Gamma =0.04$, $h =0.05$]{\includegraphics[scale=0.4]{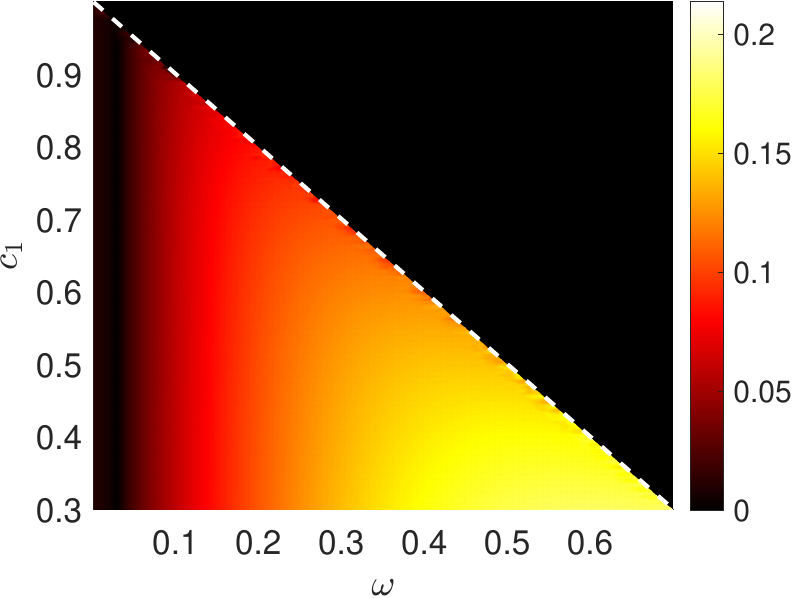}\label{heatmap2}}\,\,
	\subfloat[$\Gamma =0.04$, $h =-0.05$]{\includegraphics[scale=0.4]{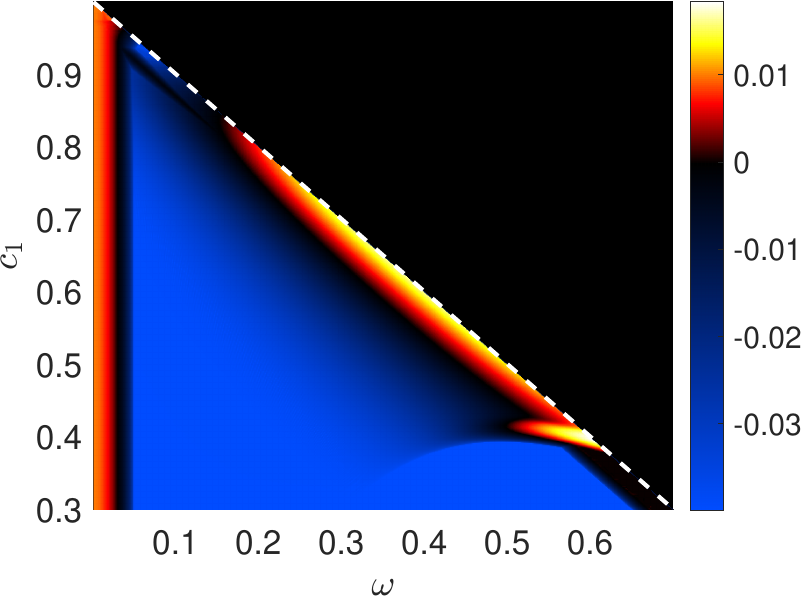}\label{heatmap3}}
	\caption{Maximum real part of the linearization eigenvalues, \(\max \operatorname{Re}(\lambda)\), in the \((\omega,c_1)\)-parameter plane. The three panels correspond to (a) \(\Gamma=0\), \(h=0\), (b) \(\Gamma=0.04\), \(h=0.05\), and (c) \(\Gamma=0.04\), \(h=-0.05\). Values at or below zero, up to numerical tolerance, correspond to spectrally stable stationary edge states, whereas positive values indicate instability. The dashed curve marks the boundary of the edge-localized existence region predicted by the analytical approximation.}
	\label{fig:heatmap}
\end{figure*}

We are also interested in the stability region of the edge states in the \((\omega,c_1)\)-parameter plane. Figure~\ref{fig:heatmap} shows the largest real part of the spectrum \(\max(\operatorname{Re}(\lambda))\). This quantity provides a direct numerical measure of stability: values at or below zero, up to numerical tolerance, correspond to spectrally stable stationary states, whereas positive values indicate instability. The dashed curve marks the boundary of the existence region predicted by the analytical approximation. The conservative case and the two driven--damped cases exhibit markedly different stability landscapes. Most importantly, the inclusion of damping and parametric forcing 
fundamentally alters the stability properties of the nonlinear edge states by generating two phase-locked dissipative branches with very different spectral characteristics.

The contrast between panels~(b) and~(c) is particularly striking. For positive driving, \(h=0.05\), the edge states remain unstable over a substantial portion of the existence region and exhibit relatively large instability growth rates. In contrast, for negative driving, \(h=-0.05\), the edge states develop considerably larger stability regions and significantly smaller instability growth rates even when unstable. This demonstrates that the combined action of damping and parametric driving, therefore, can provide a powerful mechanism for stabilizing nonlinear topological edge states. 

\subsection{Time-dependent dynamics}
\label{subsec:time_dynamics}

Finally, we compare the time evolution of selected edge-localized states in the reduced SSH amplitude system and in the original parametrically driven Klein--Gordon pendulum chain. This comparison is important because the SSH amplitude equation is obtained as an effective model for the slow dynamics of the full mechanical lattice.

\begin{figure*}[tbhp]
	\centering
	\subfloat[$\Gamma =0$ and $h =0$, $c_1=0.6$ and $\omega=0.37$]{\includegraphics[scale=0.35]{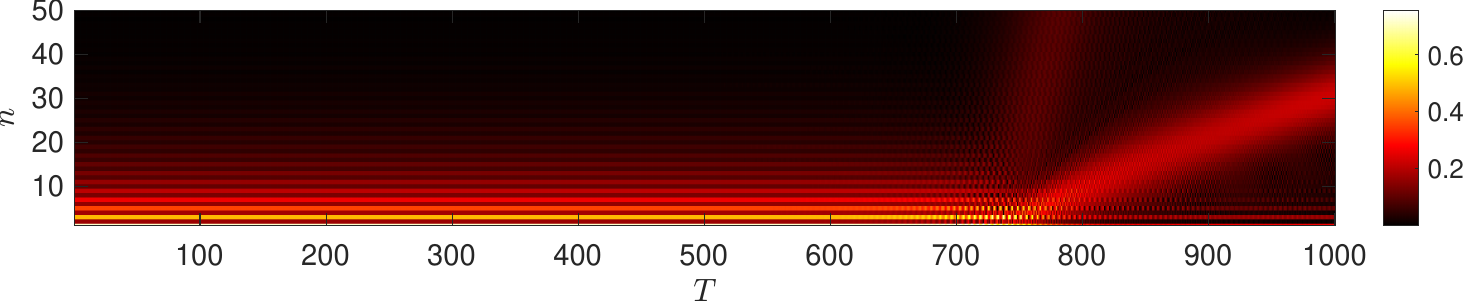}\label{ssh_dynamics_case1}}\,\,
	\subfloat[$\Gamma =0$ and $h =0$, $c_1=0.6$ and $\omega=0.37$, $\epsilon=0.02$]{\includegraphics[scale=0.35]{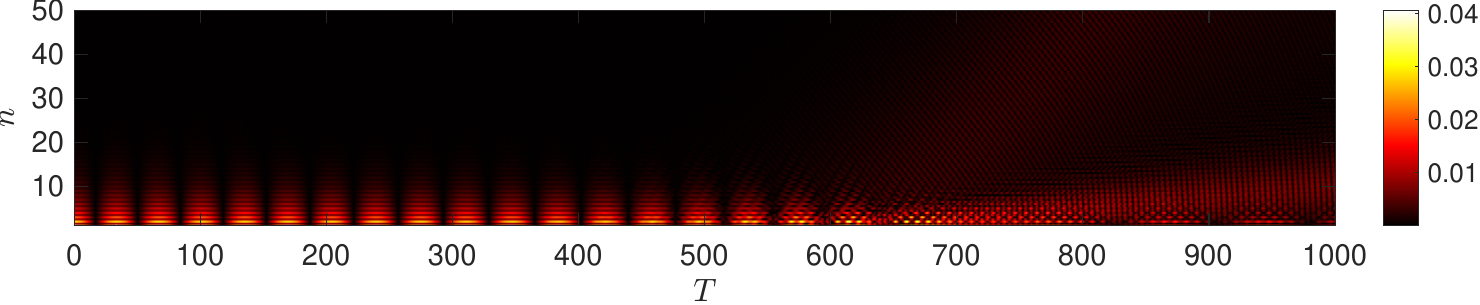}\label{kg_dynamics_case1}}\,\,
	\subfloat[$\Gamma =0.04$ and $h =0.05$, $c_1=0.6$ and $\omega=0.37$]{\includegraphics[scale=0.35]{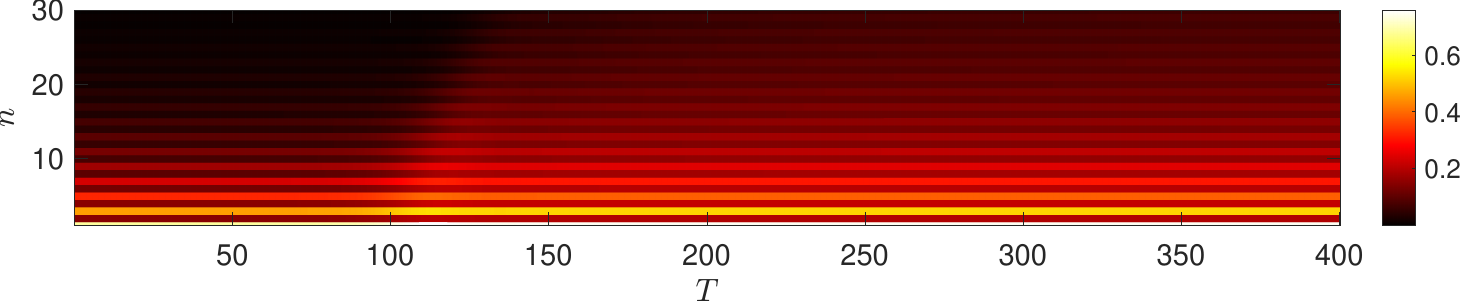}\label{ssh_dynamics_case2}}\,\,
	\subfloat[$\Gamma =0.04$ and $h =0.05$, $c_1=0.6$ and $\omega=0.37$, $\epsilon=0.02$]{\includegraphics[scale=0.35]{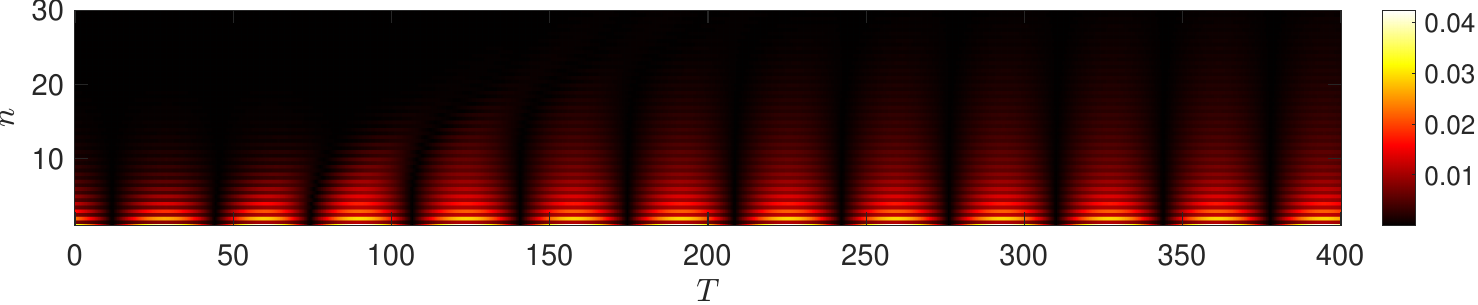}\label{kg_dynamics_case2}}\,\,
	\subfloat[$\Gamma =0.04$ and $h =-0.05$, $c_1=0.6$ and $\omega=0.37$]{\includegraphics[scale=0.35]{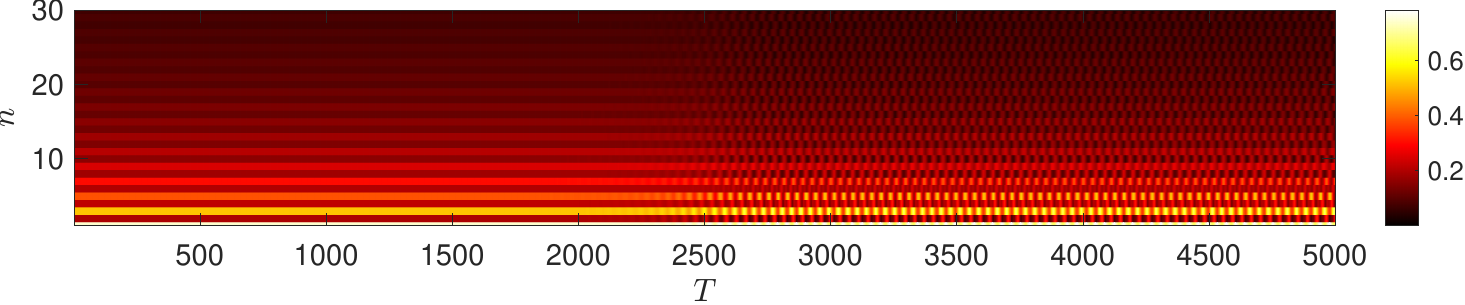}\label{ssh_dynamics_case3}}\,\,
	\subfloat[$\Gamma =0.04$ and $h =-0.05$, $c_1=0.6$ and $\omega=0.37$, $\epsilon=0.02$]{\includegraphics[scale=0.35]{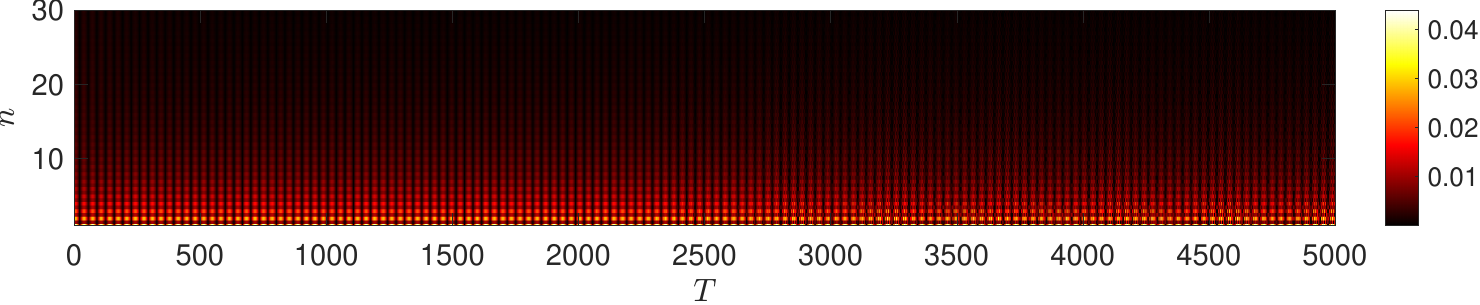}\label{kg_dynamics_case3}}
	\caption{Time evolution of representative edge-localized states in the reduced SSH amplitude system and in the original parametrically driven Klein--Gordon pendulum chain. Panels (a), (c), and (e) show the dynamics of the SSH amplitude system \eqref{eq:ssh-pddnls}, whereas panels (b), (d), and (f) show the corresponding dynamics of the Klein--Gordon chain \eqref{eq:nondim} with \(\epsilon=0.02\). The color scale represents \(|\psi_n|\) for the reduced amplitude model and \(|u_n|\) for the Klein--Gordon chain. The parameters are \(c_1=0.6\) and \(\omega=0.37\), with (a,b) \(\Gamma=0\), \(h=0\), (c,d) \(\Gamma=0.04\), \(h=0.05\), and (e,f) \(\Gamma=0.04\), \(h=-0.05\). The comparison illustrates how the reduced model captures the qualitative edge-localized dynamics of the full pendulum chain, while damping and parametric forcing modify the radiation and localization behavior.}
	\label{fig:dynamics}
\end{figure*}

Figure~\ref{fig:dynamics} shows representative simulations for the conservative case and for the two driven--damped cases. Panels (a), (c), and (e) correspond to the reduced SSH amplitude system, while panels (b), (d), and (f) show the corresponding dynamics in the original Klein--Gordon model with \(\epsilon=0.02\). The simulations demonstrate that the reduced model captures the main qualitative features of the edge-localized dynamics observed in the full system. In the Klein--Gordon simulations, small-amplitude radiation and long-time modulation are more visible, as expected from the presence of higher-order effects that are not fully retained in the amplitude equation.

The dynamical simulations further highlight the qualitative differences between the two phase-locked dissipative branches. In the conservative case and in the driven-damped case with \(h=0.05\), the instability manifests as a gradual deformation of the localized profile and enhanced radiation into the bulk. In contrast, the solution shown in panels~(e) and~(f), corresponding to \(h=-0.05\), displays remarkably robust edge localization.

This observation is particularly noteworthy because the corresponding stationary state is not spectrally stable. As shown in Fig.~\ref{fig:prof}(c), the solution possesses a pair of unstable eigenvalues with positive real part. Nevertheless, the instability growth rate is sufficiently small that the localized wave undergoes almost no visible deformation throughout the entire simulation interval. The profile remains strongly concentrated near the edge, and no significant spreading into the bulk is observed within the computational domain.

From a practical perspective, the solution therefore behaves as a long-lived localized state despite its formal spectral instability. This robustness is a direct consequence of the interplay between damping and parametric driving at the appropriate phase relation. The results indicate that the sign of the drive can have a profound effect on the persistence of nonlinear topological localization: one phase-locked branch remains relatively fragile, whereas the other supports edge states that are stable or effectively stable over experimentally relevant time scales.

Overall, the numerical results confirm the validity of the analytical approximation near onset, reveal the continuation of the nonlinear edge-state branches away from threshold, and demonstrate that parametric driving and damping can substantially enhance the robustness of nonlinear topological edge solitons. The direct simulations further support the role of the reduced SSH amplitude equation as an effective model for the slow dynamics of parametrically driven edge states and illustrate how the appropriate driven--dissipative balance can generate long-lived localized excitations even in parameter regimes where spectral instability persists.

\section{Conclusion}
\label{sec:conclusion}

We have investigated topological edge solitons in a parametrically driven and damped nonlinear SSH lattice. Starting from a vertically driven pendulum chain, we derived an effective driven--damped discrete nonlinear Schr\"odinger equation with alternating couplings and identified the conditions under which stationary edge-localized states emerge.

The linear analysis revealed the spectral structure of the system and provided the threshold for the onset of nonlinear edge branches. Near resonance, this threshold is determined by the balance between parametric driving and damping. To describe the nonlinear regime, we developed an analytical approximation based on an exponentially localized ansatz. The approximation captures the spatial decay and power--detuning relation of the edge states and shows good agreement with numerical computations near onset.

A central finding of this work is that parametric driving and damping fundamentally alter the structure and stability of the nonlinear edge states. In contrast to the conservative SSH lattice, the driven--damped system supports two phase-locked dissipative edge-soliton families corresponding to different phase relations with the external drive. These families possess markedly different stability characteristics. One branch remains predominantly unstable, whereas the other develops substantially larger stability regions and significantly smaller instability growth rates.

The numerical simulations further reveal that spectral instability does not necessarily imply a rapid loss of localization. In particular, the robust dissipative branch supports edge states that remain strongly localized over very long time intervals even when weakly unstable. Direct simulations of both the reduced SSH model and the original driven pendulum chain confirm the persistence of these localized excitations and demonstrate the validity of the reduced description.

These results identify parametric driving and damping as an effective mechanism for controlling nonlinear topological localization. By appropriately tuning the driven--dissipative balance, it is possible to substantially enhance the robustness and persistence of edge solitons. The present work therefore provides a route toward the realization and control of long-lived nonlinear topological excitations in active mechanical, optical, and metamaterial lattices.

\section*{Acknowledgement}
R.R.\ is funded by the Directorate of Research and Development, Universitas Indonesia, under Hibah PUTI Q2 2025 (Grant No.\ PKS-369/UN2.RST/HKP.05.00/2025).  R.K. acknowledges that this research is funded by the ITB Research Program 2026 under the ITB International Research Scheme through the Directorate of Research and Innovation, Institut Teknologi Bandung (Project ID: FMIPA.PN-6-126-2026). HS acknowledges support by Khalifa University through a Research \& Innovation Grant under project ID KU-INT-RIG-2024-8474000789. 
	
\bibliographystyle{apsrev4-2}
\bibliography{references}
	
\end{document}